\documentclass[prd,superscriptaddress,showpacs]{revtex4}
\usepackage{mathrsfs}
\usepackage{amssymb}
\newcommand{\D}{{\mathrm{D}}}

\newcommand{\be}{\begin{equation}}
\newcommand{\ee}{\end{equation}} 
\newcommand{\eei}{\end{equation}\indent\indent}
\newcommand{\bc}{\begin{center}}
\newcommand{\ec}{\end{center}}
\newcommand{\ber}{\begin{eqnarray}}
\newcommand{\eer}{\end{eqnarray}}
\newcommand{\ba}{\begin{array}}
\newcommand{\ea}{\end{array}}

\newcommand{\bb}{{\cal B}}

\newcommand{\ii}{{\cal I}}
\newcommand{\ff}{{\cal F}}
\newcommand{\bbeta}{{\cal \beta}}

\newcommand{\sfrac}[2]{{\textstyle{#1\over#2}}}
\def\case#1/#2{\textstyle\frac{#1}{#2} }
\newcommand{\bra}[1]{\left(#1\right)}
\newcommand{\bras}[1]{\left[#1\right]}
\newcommand{\brac}[1]{\left\{#1\right\}}
\newcommand{\curl}{{\mathsf{curl}\,}}
\newcommand{\di}{{\mathsf{div}\,}}
\newcommand{\reff}[1]{(\ref{#1})}

\begin{document}

\title{On inhomogeneous magnetic seed fields and gravitational waves
within the MHD limit}

\author{Caroline Zunckel}
\email{clz@astro.ox.ac.uk} \affiliation{Astrophysics, University of
Oxford, Denys Wilkinson Building, Keble Road, Oxford OX1 3RH, UK}
\affiliation{Department of Mathematics and Applied Mathematics,
    University of Cape Town, 7701 Rondebosch, South Africa}
\author{Gerold Betschart}
\email{gerold@phys.huji.ac.il} \affiliation{Department of
Mathematics and Applied Mathematics,
    University of Cape Town, 7701 Rondebosch, South Africa}
\affiliation{Racah Institute of Physics, Hebrew University of
Jerusalem, Givat Ram, 91904 Jerusalem, Israel}
\author{Peter K S Dunsby}
\email{pksd@maths.uct.ac.za}
\affiliation{Department of Mathematics and Applied Mathematics,
    University of Cape Town, 7701 Rondebosch, South Africa}
\affiliation{South African Astronomical Observatory, Observatory
7925, Cape Town, South Africa}
\author{Mattias Marklund}
\email{mattias.marklund@physics.umu.se}
\affiliation{Department of Physics, Ume{\aa} University, SE-901 87
Ume{\aa}, Sweden}
\begin{abstract}
In this paper we apply second-order gauge-invariant perturbation
theory to investigate the possibility that the non-linear coupling 
between gravitational waves (GW) and a large scale \emph{inhomogeneous} 
magnetic field acts as an amplification mechanism in an `almost'
Friedmann-Lema{\^i}tre-Robertson-Walker (FLRW) Universe. The spatial
inhomogeneities in the magnetic field are consistently implemented
using the {\em magnetohydrodynamic} (MHD) approximation, which yields an
additional source term due to the interaction of the magnetic field
with velocity perturbations in the plasma. Comparing the solutions
with the corresponding results in our previous work indicates that,
on super-horizon scales, the interaction with the spatially
inhomogeneous field in the dust regime induces the same boost as
the case of a homogeneous field, at least in the ideal MHD
approximation. This is attributed to the observation that the MHD
induced part of the generated field effectively only contributes on
scales where the coherence length of the initial field is less than
the Hubble scale. At sub-horizon scales, the GW induced magnetic
field is completely negligible in relation to the MHD induced field.
Moreover, there is no amplification found in the long-wavelength
limit.
\end{abstract}
\pacs{98.80Cq}

\maketitle
\section{Introduction}
Astrophysical observations indicate that almost all environments in
the Universe are magnetised, and the more we search for
extragalactic fields, the more pervading they are revealed to be.
Such cosmological magnetic fields are found in galaxy clusters, disk
and spiral galaxies as well as in high-redshift condensations (see
\cite{grasso-widrow} for reviews). A fascinating and as yet unsolved
question is how did these fields originate? The current properties
of magnetic fields should, in principle, reflect their past and give
clues to their origins, so we rely on observations to contribute to
finding an answer to this important question.

Faraday rotation and Zeeman splitting measurements indicate that
galactic magnetic fields at high redshifts exist with roughly the
same strength, $10^{-7}$ to $10^{-5}\,\mathrm{G}$, as those found in
the Milky Way \cite{Kronberg,Zweibel-Heiles}. The common properties
of large scale fields in different galaxies indicates that their
origins may be intrinsically connected to the cosmological
repercussions of the interplay between gravitational and gauge
interactions \cite{D}. This suggests that their origin may be
primordial, in which case their presence could be related to the
physics of the very early Universe. One example is Big Bang
Nucleosynthesis, where the interaction of a magnetic field with the
magnetic moment of a neutrino may have given rise to a spin-flip and
change of its handedness, introducing an additional neutrino degree
of freedom \cite{enqvist}.

Depending on their spectrum, magnetic fields existing in
proto-galactic clouds with strengths of $10^{-12}$ to
$10^{-9}\,{\mathrm G}$ may have played a significant role in
structure formation \cite{wasserman}. The Lorentz force that acts on
charges in an inhomogeneous (i.e. $\curl B\neq 0$) magnetic field
has been shown to induce peculiar velocities \cite{cheng} which seed
density perturbations, and in so doing alter the gravitational
instability picture. Moreover,  hyper-magnetic fields, hypothesised
to emerge during the electroweak phase, have also been identified as
a possible source of the observed baryon asymmetry of the Universe
\cite{mass-giov}. For these reasons, the determination of the origin
and properties of cosmic magnetic fields is of extreme importance in
cosmology. This makes magnetogenesis, the determination of a self-consistent 
theory for the generation of cosmological magnetic fields with the
strengths and on the scales measured today, one of the `hot' topics
in modern cosmology. The most popular theories include the
amplification of a small field by the galactic dynamo and the
adiabatic proto-galactic collapse at the start of structure
formation. Although these mechanisms are shown to yield
substantial enhancement, they are not self-sufficient as they
presuppose the existence of seed fields. In addition, these seed
fields must satisfy very stringent strength and size criteria
in order for the generated fields to agree with the magnitudes
observed today. It follows that the problem we face is to provide a
mechanism that induces a large enough amplification of a
weak \emph{pre-existing} seed field, so that the aforementioned
mechanisms are physically viable.

In this paper, we extend the work of Betschart \emph{et al.}
\cite{us}, which investigated the coupling between a large-scale
{\em homogeneous} magnetic field and the gravitational wave spectrum
which accompanies most inflationary scenarios. This work built on
earlier work by Tsagas {\it et al.} \cite{GWamp} in which the same
interaction was studied within the weak field approximation
\cite{tsagas, TB}. The analysis in \cite{us} demonstrated that this
coupling can lead to an amplification, provided the dimensionless 
shear anisotropy $\sigma/H$ at the end of inflation is larger 
than $10^{-40}$.

In this investigation, we consider the general case where
the original magnetic field is \emph{inhomogeneous} over a
typically observed coherence scale. By comparing this analysis
with our treatment of homogeneous fields in \cite{us}, we aim
to determine the implications of placing restrictions
(such as homogeneity) on the properties of primordial magnetic
seed fields.

The highly non-linear nature of the Einstein's field equations makes
finding exact solutions as well as applying numerical techniques
extremely difficult. In order to solve them analytically, severe
symmetry assumptions are often required to simplify the physical
models, which then restricts their applicability. For cosmological
applications, using a perturbative approach, which entails
decomposing the real physical Universe into a family of spacetimes,
yields surprisingly good results. This allows us to encode the
inhomogeneities we see today as perturbations expanded
around a fictitious idealised background model, most commonly taken 
to be the Friedmann-Lema{\^i}tre-Robertson-Walker (FLRW) spacetimes.
Before we can implement perturbation theory in a self-consistent way,
we must give due attention to the issue of gauge-invariance which has
historically plagued such studies \cite{harmonics}.

If we consider a weak large-scale magnetic field residing in a
background FLRW model as done by Tsagas {\it et al.} \cite{GWamp},
the interaction with linearised gravitational waves manifests itself
as a first-order perturbation of this background. The problem with
this approach is that it is not gauge-invariant in a strict
mathematical sense, due to the fact that the magnetic field
introduces a preferred direction and therefore breaks the isotropy
of the background FLRW model. This problem is partially overcome by
assuming that the magnetic field is weak and that its contribution
to the energy-momentum tensor is such that it does not disturb the
isotropy of the FLRW background \cite{tsagas}.

A completely self-consistent solution to this problem is obtained by
treating any seed magnetic fields as a first order perturbation
and including the interaction with gravitational perturbations by
going to second order in perturbation theory \cite{us}.

The introduction of magnetic spatial gradients at linear order
requires a more subtle treatment of the associated spatial currents,
requiring the use of the magnetohydrodynamic approximation (MHD) to
provide a framework in order to obtain a tractable solution. This
single-component fluid model provides an accurate description of a
two-species plasma when effects occurring over much larger time and 
length scales than those characteristic of plasma effects are studied. 
This allows one to handle low-frequency phenomena in a magnetised plasma 
using the standard machinery of fluid dynamics. This reduced description is
achieved by defining appropriate one-fluid variables representing
the bulk quantities, while Ohm's law provides a consistent treatment
of the associated electric field. It is common practice to take
advantage of the high conductivity, $\sigma$, of the young cosmic
plasma and employ the ideal MHD limit $(\sigma\rightarrow\infty)$.
In this limit, the flux lines are effectively glued to the plasma
fluid elements. In the non-ideal MHD limit \cite{marklund}, where
$\sigma$ is assumed to be large but not infinite, the electric field
enters at linear order in the case of a first-order magnetic field.
Here we focus on the \emph{ideal} MHD case which is shown to be
equivalent to assuming that the observed electric field vanishes in
the rest frame of the fluid.

The mathematical framework we use is the 1+3 covariant
approach \cite{covariant,EE,cargese} to perturbation theory
which allows Maxwell's  and Einstein's equations to be written
in an intuitive and simple fashion \cite{marklund}. The covariant
definition of the variables ensures that their connection to
physically and geometrically significant quantities is immediately
transparent and their exact presentation gives them meaning in any
spacetime.  Most importantly, the identification of gauge-invariant
(GI) perturbation variables at a given order is relatively 
straightforward.

It is found that the generated magnetic field consists of
contributions from two sources. The first stems from the interaction
between GWs and the `background' magnetic field, the second comes
from the rotation of the induced electric field, which is caused in
MHD by the `background' magnetic field interacting with the
velocity perturbations in the plasma. Comparing the solutions with
the corresponding results in \cite{us} reveals that, at
super-horizon scales, the interaction of GWs with a spatially
inhomogeneous magnetic field in the dust regime yields an
amplification of the same order of magnitude as found in the case of
a homogeneous field, at least in the ideal MHD approximation. The
MHD induced part of the magnetic field becomes important only at
sub-horizon scales, where the GW induced contribution is negligible.
In other words, the contribution stemming from the GWs dominates
over plasma effects in the ideal MHD limit at super-horizon scales,
whereas the roles are interchanged at sub-horizon scales.

The units employed in this paper are $c=h=1$ and $\kappa=8\pi G=1$.

\section{Perturbation scheme}

We employ the same perturbative scheme as in Betschart \emph{et al.}
\cite{us}. Since the presence of a magnetic vector field in the FLRW
background does not yield a strictly gauge-invariant system, we
introduce such fields as a perturbation of the FLRW background.
Similarly, the gravitational and velocity perturbations are required
to vanish in this background for them to be gauge-invariant. Since
we are interested in the interaction of magnetic fields with linear
gravitational wave distortions, we need to treat this problem at
second-order in perturbation theory. Using the 1+3 covariant
approach \cite{covariant,EE,cargese}, we expand the physically
relevant variables in terms of two smallness parameters to
distinguish between the magnitudes of the inhomogeneous magnetic
field ($\sim \epsilon_{B}$) and the amplitude of the GWs ($\sim
\epsilon_g$). It follows that the magnitude of the interaction of
interest is of order $\mathcal O(\epsilon_{B}\epsilon_{g})$. Since
we are only interested in the cross interaction of the magnetic and
gravitationally sourced perturbations, we only retain mixed terms of
order $\mathcal O(\epsilon_{B}\epsilon_{g})$ and neglect those of
order $\mathcal O(\epsilon_g^2)$ and $\mathcal O(\epsilon_{B}^2)$.
In fact, such terms always appear in the calculations concerning the
induced magnetic field, multiplied by terms of order $\mathcal
O(\epsilon_{B})$ and therefore do not lead to inconsistencies in
the perturbation scheme.

The perturbation spacetimes may be split as follows \cite{us}:
\begin{itemize}
\item{$\mathcal B$ = Exact FLRW as the background spacetime,
$\mathcal{O}(\epsilon^0)$;} \item{$\mathcal F_1$ = Exact FLRW
perturbed by an inhomogeneous magnetic field whose energy density
and anisotropic stress are neglected, $\mathcal O(\epsilon_{B})$;}
\item{$\mathcal F_2$ = Exact FLRW with gravitational and velocity
perturbations\footnote{We assume that the velocity perturbations
are sourced by gravitational fluctuations.} $\mathcal
O(\epsilon_{g})$;}
\item{$\mathcal S$ =
$\mathcal F_1+\mathcal F_2$ allows for inclusion of interactions
terms of order $\mathcal O(\epsilon_{B}\epsilon_{g})$.}
\end{itemize}
We will generally refer to terms of order  $\mathcal
O(\epsilon_{B})$ and $\mathcal{O}(\epsilon_g)$ appearing in
$\mathcal F$ as `first-order' and to the cross terms $\mathcal
O(\epsilon_{B}\epsilon_{g})$ appearing in $\mathcal S$ as
`second-order'.

As we will discuss in more detail in the next section, the
presented hierarchy of spacetimes given above is only justified if
the electric field is at least of second order. However, we will
assume the ideal MHD case, which indeed implies that the electric
field must be second order.

Before we turn to the MHD approximation, we first describe the
basic equations describing the background and first-order
spacetimes keeping the equations as general as possible.

\subsection{FLRW background}

In order to perform an 1+3 decomposition of any spacetime, we need
to introduce a universal reference 4-velocity field $u^a$ relative
to which all motion is defined and quantified. In accordance
with the observed average recession of the galaxies, we assume
that the matter in the Universe has a locally well-defined
preferred motion that can be represented by a unique 4-velocity
vector field $u^a$, satisfying $u_a u^a=-1$. Based on the
Copernican principle, we can assume that this holds at each point
in the Universe. We then introduce a family of observers, called
the {\em fundamental observers}, travelling such that this field
represents the congruence of their worldlines. In so doing, any
observations made are those relative to this preferred frame.
In choosing this field to coincide with the average velocity of
the matter in the Universe, it acquires an `invariant significance'
such that the covariant quantities at every point, which are defined
with respect to $u^a$, can be decomposed uniquely \cite{cargese}.

FLRW models are spacetimes that are spatially isotropic and
homogeneous about every point. Relative to the congruence of
fundamental observers with 4-velocity $u^a$, the kinematics are
assumed to be locally isotropic. This implies that all tensorial
quantities such as the acceleration vector $\dot u_a \equiv
u^b\,\nabla_b\, u_a$, the shear $\sigma_{ab} \equiv
\D_{<a}\,u_{b>}$ and the vorticity $\omega_{ab} \equiv
\D_{[a}\,u_{b]}$ must vanish to eliminate preferred direction in
the spatial sections:
\be 0=\dot u_a=\sigma_{ab}=\omega_{ab},
\ee
The constraints
\be
\pi_{ab}=q_{a}=0\;,
\label{FLWpi}
\ee
ensure that the energy-momentum tensor and thus the Ricci tensor are
isotropic and indicate that a perfect fluid is a necessary
requirement of this model. These restrictions mean that the
electric and magnetic components of the Weyl tensor and hence the
Weyl tensor itself, vanish identically
\be
E_{ab}=H_{ab}=0\Rightarrow C_{abcd}=0\ .
\ee
We can then infer that these models are conformally flat. 
Furthermore the spatial uniformity forces the spatial gradients 
of the energy density $\mu$, the pressure $p$ and the 
expansion $\Theta$ to vanish
\be
0=\D_a\,\mu=\D_a\,\Theta=\D_a\,p . \label{FLWk2}
\ee
As usual, the spatial derivative $\D_a \equiv h_a^{~b}\,\nabla_b$ 
is obtained by projection of the spacetime covariant derivative 
$\nabla_a$ onto the 3-space (with metric $h_{ab} \equiv g_{ab}+u_au_b$) 
orthogonal to the observer's worldline. As a consequence, the key background
equations are the energy conservation equation
\be
\dot{\mu}+\Theta(\mu+p)=0\ ,
\label{energy}
\ee
the Raychaudhuri equation
\be
\dot{\Theta}=-\sfrac{1}{3}\Theta^2-\sfrac{1}{2}\bra{\mu+3p}+\Lambda,
\label{Raychaudhuri}
\ee
and the Friedmann equation
\be
\mu+\Lambda=\frac{1}{3}\Theta^2+\frac{3K}{a^2},\label{Friedmann}
\ee
describing the intrinsic curvature of the homogeneous and isotropic 3-spaces. 
The curvature constant $K$ indicates the geometry of the Universe and 
can be normalised to $K=-1, +1, 0$ for spatially open, closed and 
flat Universes.
\subsection{First-order perturbations}\label{sec:inhomofirst}
\subsubsection{The inhomogeneous magnetic field $\tilde B_a$}
We assume that the seed magnetic field $\tilde B_a$ residing in
the ${\mathcal F}_{1}$ spacetime is inhomogeneous over its typical
coherence scale. The spatial gradients $D_b \tilde B_a$ are thus
of order ${\mathcal O}(\epsilon_B)$. Given that the magnetic field
is a first-order perturbation on the background, the magnetic
anisotropy $\pi_{ab}=-\tilde B_{<a}\tilde B_{b>} \sim {\mathcal
O}(\epsilon_{B}^2)$ can be neglected~\footnote{Here the angle bracket
represents the projected symmetric trace-free (PSTF) part of any
tensor: $A_{<ab>}\equiv
h^c{}_{(a}h^d{}_{b)}A_{cd}-\sfrac{1}{3}h_{ab}A^c{}_c$.}. Since the
associated electric field is perturbatively smaller than the
magnetic field and enters only in ${\mathcal S}$ as will be argued
in the next section, the magnetic induction equation  has the form
\be \label{ind}
\dot{\tilde
B}_{<a>}+\sfrac{2}{3}\Theta\tilde B_{a}=0 .
\ee
Thus the magnetic field decays as \be \tilde B_a=\tilde
B^0_a\bra{\frac{a_0}{a}}^2 ,\label{Bscale} \ee where $a$ denotes
the scale factor, i.e., $\Theta = 3\, \dot a/a = 3 H$ and 
$H$ denotes the inverse Hubble length. The adiabatic decay evident
in equation \reff{Bscale} arises from the expansion of the
Universe which conformally dilutes the field lines due to flux
conservation \cite{kahniashvili}. By taking the spatial gradient
of equation \reff{ind}, it is easy to show that the gradient of
the magnetic field evolves as $D_b \tilde B_a \sim a^{-3}$. Note
also that the induction equation does not discriminate between
homogeneous and inhomogeneous magnetic fields.
\subsubsection{Gravitational waves}
In the covariant approach to cosmology, linearised gravitational
waves are purely tensorial and are monitored via the electric
$(E_{ab})$ and magnetic $(H_{ab})$ Weyl constituents, which are not
sourced by rotational (vector) and density (scalar) perturbations
\cite{chal, GW}. The transverse (divergence-free) nature of these
projected, symmetric, trace-free (PSTF) tensors means that we only
need to eliminate their vector parts in order for them to
characterise frame-invariant GWs. We isolate the linear tensorial
modes by imposing the constraints
\be 0=\D_a\mu=\D_a p=\D_a \Theta=\omega_a=\dot{u}_{a}
.\label{GWconditionshomo} \ee These restrictions ensure that the
sources of vector modes (spatial gradients and vector perturbations
themselves) vanish and lead to the constraints~\footnote{We use
$\curl V_a \equiv \epsilon_{abc}\,D^b\,V^c$ to denote the $\curl$ of
a vector and $\curl W_{ab} \equiv
\epsilon_{cd<a}\,\D^c\,W_{b>}^{~~d}$ to denote the covariant $\curl$
of a second-rank PSTF tensor, where $\epsilon_{abc}$ is the volume
element of the 3-space. Finally, the covariant spatial Laplacian is
$\D^2 \equiv \D^a\D_a$.}
\be
0=\D^a\sigma_{ab}=\D^aE_{ab}=\D^aH_{ab}=H_{ab}-\curl\sigma_{ab}
.\ee Since the shear tensor is coupled to $H_{ab}$ and $E_{ab}$,
it can also be used as a measure of gravitational waves.  The
tensorial gravitational waves are governed completely by a closed
wave equation for the shear. At linear order, this wave equation
is given by~\cite{us}
\be
\ddot{\sigma}_{<ab>}-\D^2 \sigma_{ab}
+\sfrac{5}{3}\Theta\dot{\sigma}_{<ab>}+\bra{\sfrac19\Theta^2+\sfrac{1}{6
}\mu
-\sfrac{3}{2}p+\sfrac{5}{3}\Lambda} \sigma_{ab}=0
.\label{sigmadot}
\ee
Observe that in general the RHS of equation~\reff{sigmadot} is
nonzero (see~\cite{Ma97} for the case of irrotational dust
spacetimes).
\section{Basics of ideal MHD}
Since the mean free path between the electron-ion collisions in a
typical plasma can be short compared to the characteristic MHD
length scale, it is not always obvious that a fluid description is
indeed valid. On small scales these interactions are frequent and
cause the two species to move relative to each other, generating
charge separation effects referred to as plasma oscillations. If
we consider much larger scales on which the individual collisions
are not explicitly seen (i.e. low-frequency phenomena), the
different species are observed to move together with a common
average velocity, allowing an effective single fluid description
of this two-component system. If, in addition, the characteristic
MHD length scale is much greater than the plasma Debye length and
the gyro radius, then MHD gives an accurate description of
low-frequency phenomena in a magnetised plasma (see, for example,
Refs.~\cite{marklund,MHD}).

We now pay attention to how the currents, which are established due
to the net motion of the fluid by induction, modify the field and in
so doing couple the hydrodynamical equations to Maxwell's equations
via Ohm's law.

We perform the calculations in an irrotational Universe with
vanishing cosmological constant $\Lambda$, and assume $p=0$ from
now on since it has been argued that the reduced MHD description
of a plasma is only valid in the cold limit \cite{khanna} (i.e. $p
= 0$ and non-relativistic motion of the plasma particles). These
assumptions simplify the analysis tremendously while still
allowing us to make a meaningful comparison with the results
obtained in the homogeneous case for dust \cite{us}.

As a further simplification, we adopt the \emph{geodesic} frame,
in which the acceleration of the fluid frame $\dot{u}^a$ vanishes
to all orders.
\subsection{Covariant Theory}\label{sec:phase}
We now turn to the evolution equations of magnetohydrodynamic
variables. We assume charge quasi-neutrality of the plasma (i.e. the
number densities of the electrons, $n_e$, and ions, $n_i$, are
roughly equal such that the total charge $\rho_c$ effectively
vanishes, $\rho_c=-e(n_e-n_i)\approx 0$). In the spirit of the MHD
approximation, we choose to formulate the equations using the
magnetic field as our primary variable and use Maxwell's equations
to express the electric field and currents in terms of $B^a$. In
order to identify the perturbative order of the terms which we can
eliminate without excluding physically important effects, the exact
equations that are necessary to fully describe this system are
stated.
\subsubsection{Maxwell's equations}
Under the assumption of charge neutrality and vanishing vorticity,
the behaviour of electromagnetic fields and currents in curved
spacetimes is governed by Maxwell´s equations in the following
form  (see, e.g., \cite{EE,mark,marklund}):
\ber
\dot{B}_{<a>}+\sfrac{2}{3}\Theta B_{a} &=&\sigma_{ab}B^b-\curl E_a,
\label{eq:dotB}\\
\dot{E}_{<a>}+\sfrac{2}{3}\Theta E_{a} &=&\curl B_a-j_{<a>}
+\sigma_{ab}E^b, \label{eq:dotE}\\
\D^a B_a&=&0\;,\\
\D^a E_{a}&=&0\;,
\eer
where the last two constraint equations are supposed to hold at
all orders.
\subsubsection{Equations of motion}
We assume that the interactions of the ions and electrons
collectively isotropise their motions, such that in a chosen frame
the properties of the fluid on macroscopic scales can be described
in terms of an average velocity $v^a$. We take this mean motion to
be the center of mass velocity of the electron-ion system,
defined by
\be
v^a=\frac{\mu_e v^a_{e}+\mu_i v^a_{i}}{\mu_e+\mu_i}\;, \label{v}
\ee
where $v^a$ coincides with the velocity of the fundamental
observer at zeroth order, that is, $v^a=0$ relative to the
fundamental observer at lowest order. At higher orders, the
behaviour of the electromagnetic fields also depends on this bulk
velocity.

Ohm's law is generally formulated in the local rest frame of the
conducting fluid and is assumed to hold at all orders:
\be
j_{<a>}=\sigma(E_{a}+\epsilon_{abc}v^b B^c)\;, \label{ohms}
\ee
where the conductivity $\sigma$ is taken to be a constant for
simplicity. The 3-vector $E^a$ represents the field as observed
from the rest-space of the fluid and the second term in
\reff{ohms} is the apparent electric field associated with the
bulk velocity $v^a$. We make the standard assumption that the
cosmic medium is infinitely conducting, and consequently apply the
ideal MHD limit. This is valid considering the early epoch in
which the interaction takes place and is also consistent with the
treatment of a homogeneous magnetic field in a \emph{spatially
flat} Universe in Betschart \emph{et al.} \cite{us}; the
gravito-magnetic interaction was shown in section (IV.B) of
\cite{us} to generate the same magnetic field irrespective of
conductivity. By letting $\sigma\rightarrow \infty$ in Ohm's law
\reff{ohms}, we find that $E_{a}+\epsilon_{abc}v^b B^c$ must
tend towards zero in order for the spatial current
$j^{<a>}=\rho_ev_e^a + \rho_iv_i^a$ to remain finite. The electric
field is now determined jointly by the fluid velocity and the
magnetic field:
\be
E_a=-\epsilon_{abc}v^b B^c\;. 
\label{Einhomo}
\ee
It follows that the electric field is at least of second order and
vanishes at lower orders. At first order, the magnetic field 
$\tilde B^a$ determines the current via $\curl \tilde B^a = j^{<a>}$, 
while at second order the current follows from equations~\reff{eq:dotE} 
and \reff{Einhomo}. Using charge neutrality, the evolution of the 
total energy density ${\cal E} \equiv \mu_e+\mu_i$ and the center 
of mass velocity follow from the corresponding total energy 
and total momentum conservation equations \cite{mark,marklund}:
\ber
\dot{\cal E} +  \Theta{\cal E}  &=& -\D_a\bra{{\cal E} v^a},\\
{\cal E}\bra{\dot{v}_{<a>}+\sfrac{1}{3}\Theta v_{a}} &=& -{\cal
E}\bra{v^b \D_bv_a+\sigma_{ab}v^b}-\epsilon_{abc}j^b B^c\;,
\label{eq:v}
\eer
which hold in the cold plasma limit in the geodesic frame.
\section{Second-order perturbations: The interaction}
We look to Maxwell's equation to determine the nature of the
interaction between GWs and the first-order magnetic field $\tilde
B^a$. If the back-reaction of the induced field with the shear and
the center of mass velocity is ignored, the induction equation
takes the form
\be
\dot{B}_{<a>}+\sfrac{2}{3}\Theta B_{a} =\sigma_{ab}\tilde B^b +
2\D^b\bra{v_{[a}\tilde B_{b]}}\;, \label{dotB}
\ee
where the second term on the RHS $-\curl E_a$ describes
the dragging of the field lines by the fluid. We are faced with
the problem of removing the primary magnetic field component of
$B^a$ from the LHS of \reff{dotB} to ensure that it is truly
second-order. Given that the magnetic spatial gradients are now
retained in ${\mathcal F}_1$, the commutation relation used as an
example in the homogeneous case in Betschart \emph{et al.}
\cite{us}, is now consistently satisfied when the power series
expansion of $B^a$,
\ber
B^a=\epsilon_{B} B^a_{1}+ \epsilon_g\epsilon_{B}
B^a_{2}+{\mathcal O}(\epsilon^2_g,\epsilon^2_{B})\;,
\nonumber
\eer
is applied. The reason why an inconsistency arose previously seems
to stem from the requirement that the first-order magnetic field is
homogeneous, which messed up the commutation relations. A similar
inconsistency was encountered when studying the gravito-magnetic
interaction in the vicinity of a Schwarzschild black hole, where
staticity was imposed upon the first-order magnetic field
(see~\cite{clarkson} for details). It is interesting to remark,
however, that constraining the magnetic field to be solenoidal
does not pose any problems whatsoever.

Although the above expansion does not immediately appear to be
invalid, in the event that an inconsistency might exist, we choose
to represent the magnetic field using the second-order gauge
invariant (SOGI) variable
\be
\beta_a\equiv \dot{B}_{<a>} +\sfrac{2}{3}\Theta B_a\;,
\ee
first identified in Betschart \emph{et al.} \cite{us}. We select
the interaction variable $I_a=\sigma_{ab} \tilde B^b$ and define
the variable $F_a\equiv -\curl E_a=2D^b\bra{v_{[a}\tilde B_{b]}}$.
We note that the electric field is of the order ${\mathcal
O}(\epsilon_{B}\epsilon_{g})$ and thus enters the ${\mathcal S}$
spacetime. We can now restate Maxwell's equations in ${\mathcal
S}$ as a system of differential equations in terms of these SOGI
variables
\ber
\beta_a &=& I_a + F_a\;, \label{dotBeta}\\
\D^a E_a &=& 0\;, \label{divEinhomo}\\
\D^a{B}_{a} &=& 0\;. \label{divBinhomo}
\eer
To close the system, we use the velocity propagation equation
\reff{eq:v}. Given that both $v^a$ and $\tilde B^a$ are
individually regarded as first-order, only the linear
part of this equation is needed:
\be
\dot{v}_{<a>}+\sfrac13\Theta v_a=0\;.
\label{dotv}
\ee
 From equation \reff{dotBeta}, we see that the generated magnetic
field $B^a$ can be found directly by integrating a linear
combination of the $I^a$ and $F^a$ solutions once they are found.

It is convenient to re-scale the magnetic field variable by
defining $\bb_a=B_a\bra{a/a_0}^2$. The time dependence of $\bb^a$
found in the final solutions then describes the evolution of the
\emph{generated} field relative to the `background' field. The
main variables become
\be
\bbeta_{a}=\bra{\frac{a_0}{a}}^2\dot{\bb}_{<a>}\;,
\quad~~~\ii_{a}=\sigma_{ab}\tilde
\bb^b\;,\quad~~~~ \ff_a=2 \D^b\bra{v_{[a}\tilde\bb_{b]}}\;.
\ee
Using $H=\Theta/3=\dot a/a$ we can restate the
important equation \reff{dotBeta} in terms of
these variables as
\be
\dot{\bb}_{<a>}= \ii_a +\ff_{a}\;. \label{beta2}
\ee
Integrating this equation with respect to proper time yields
$\bb^a$; the constant of integration is determined by the
physical condition that at the time $t_0$, when the interaction
begins, there is no generated magnetic field, so we have  
$\bb^a_0=B^a_0=0$ initially.
\section{Evolution equations for the main variables}
It is evident from equation \reff{beta2} that the SOGI magnetic
field $\bb^a$ can be extracted directly by integrating a linear
combination of the solutions for the interaction $\ii^a$ and the
electric field rotation $\ff^a$. We turn to find the evolution
equations for these variables, maintaining generality.
\subsection{Harmonics}
We employ the standard harmonic decomposition \cite{harmonics,harrison}
to deal with the Laplacian operator present in the wave
equation of the shear. It is standard procedure to assume that the
time and spatial dependence of each variable is separable, so
that the variable can be expressed as the product of the time 
and spatial parts. This operation effectively decomposes the
differential equation for the time evolution of a perturbation 
variable into separate equations describing the time evolution
of each harmonic component which characterised by a 
comoving wavenumber $k$. Since any perturbation of a quantity can be 
expressed as the superposition of normal modes, we can decompose 
the spatial part into the summation over a series of
harmonics $Q^{(k)}$ which are covariantly constant
$\dot{Q}^{(k)}=0$ and are chosen to be the eigenfunctions of 
the Laplace-Beltrami operator
\be
\D^2 Q^{(k)}=-\frac{k^2}{a^2}Q^{(k)}\;. \label{Q}
\ee
We can then define a comoving scale $\lambda=2\pi a/k$ for each
perturbation associated with a harmonic function $Q^{(k)}$. In our
application, the harmonic decomposition is particularly useful as it
allows us to distinguish the specific situation where the
wavelengths of the perturbations are much larger than the Hubble
scale $(2\pi a/k\gg{\mathrm H}^{-1})$, in which case the Laplacian
operator in equation \reff{Q} that is proportional to $k^2$, can be
eliminated, yielding differential equations which are easier to 
solve. Although the use of a plane wave description is mathematically 
incorrect in curved space, it turns out that the only difference that 
arises is the allowable values of the wavenumbers. For a flat 
geometry $(K = 0)$ the eigenvalues form a continuous spectrum 
where $k^2\geq 0$. The spectrum for an open model $(K = +1)$ is 
discrete with $k^2=\alpha(\alpha+2)$ where $\alpha=1,~2,~3..$. 
A spacetime with negative curvature $(K= -1)$ can accommodate 
the eigenvalues $k^2=1+\alpha^2$ where $\alpha^2 \geq 0$ \cite{harrison}.
\subsection{Governing equation for the interaction
variable $\ii^a$}\label{sec:inhomointeraction}
In section (B.1) of Betschart \emph{et al.} \cite{us}, the
solution for $I^a$ obtained by solving its wave equation was found
to agree with the result calculated from the multiplication of the
time dependencies of the shear and background magnetic field,
determined individually. Solving the wave equation for the
interaction variable, however, requires a harmonic decomposition.
Since the interaction variable is not necessarily divergence-free,
$\D^aI_a = \sigma_{ab}\D^a\tilde B^b \neq 0$ in general, it
has a non-zero scalar contribution. This contribution is however 
exactly equal to $\D_a\beta^a$ and therefore drops out of Maxwell's 
equation \reff{eq:dotB} rendering it purely solenoidal. 
Although decomposing the interaction term $\ii^a$ as a pure vector 
(as in the homogeneous case) is therefore incorrect, it is
still possible to find the solution for $\ii^a$: either by solving
for $\sigma_{ab}$ and the first-order part of $\bb^b$ separately
and then multiplying the solutions obtained, or by proceeding in a
similar manner as in \cite{us}.

GWs are purely tensorial and so we expand the representative shear
variable with the help of tensor harmonics $Q^{(k)}_{ab}$,
\be
\sigma_{ab}=\sum_{k}\sigma^{(k)}Q^{(k)}_{ab}\;,\label{shearharm}
\ee
where as usual $\dot Q^{(k)}_{<ab>}=0$ and
$\D^2Q^{(k)}_{ab}=-(k^2/a^2)Q^{(k)}_{ab}$ hold. Each gravitational
wave mode is associated with the physical wavelength
\be
\lambda_{\mathrm{GW}}=2\pi a/k\;.
\ee
The expansion of the magnetic field in pure vector (solenoidal)
harmonics is
\be \tilde B_{a}=\sum_{n}\tilde B^{(n)}Q^{(n)}_{a}\;,
\ee
and these vector harmonics obey the relations $\dot
Q^{(n)}_{<a>}=0$ and $\D^2Q^{(n)}_{a}=-(n^2/a^2)Q^{(n)}_{a}$.
Similarly as above, we associate with a given wavenumber $n$
characterising a magnetic perturbation a characteristic length
scale,
\be
\lambda_{\tilde B} =2\pi a/n\;,
\ee
which we relate to the size of the magnetised region.

To simplify the treatment of the interaction between GWs and the
magnetic field, we proceed as follows. First, we assume the
magnetised plasma region has a finite size $\lambda_{\tilde
B}=2\pi a/m$ corresponding to some wavenumber $m$, which encodes
the magnetic inhomogeneity over this region. We therefore write
$\tilde B_{a}=\sum_{n\geq m>0}\tilde B^{(n)}Q^{(n)}_{a} \approx
\tilde B^{(m)}Q^{(m)}_{a}$. Since the interaction is most
effective if the gravitational wavelength $\lambda_{\mathrm{GW}}$
matches the size of the magnetic field region $\lambda_{\tilde
B}$~\cite{us}, we may restrict ourselves to the resonant case
where the gravitational and magnetic wavenumbers agree. This means
that the comoving scale of the magnetic field perturbation is the
same as that of the GWs, i.e. $k=m$. Consequently, the
$m$-mode of the shear is the main contribution to the interaction,
which now reduces to a single mode-mode term: $\ii_a \approx
\sigma^{(m)}\tilde\bb^{(m)}Q_{ab}^{(m)}Q^b_{(m)}$. Making use of
these considerations, it is now a straightforward task to obtain a
closed equation for the interaction variable $\ii_a$ by combining,
as in~\cite{us}, the standard evolution equations for the shear and
the electric Weyl tensor, and using $\dot{\tilde B}_{<a>}=0$. In this 
way we readily arrive at
\be
\ddot{\ii}_{<a>}+ \frac{m^2}{a^2} \ii_a
+\sfrac53\Theta\dot{\ii}_{<a>}+
\bra{\sfrac{1}{9}\Theta^2+\sfrac{1}{6}\mu-
\sfrac{3}{2}p+\sfrac{5}{3}\Lambda}\ii_a=0\;. \label{eq:ddotI}
\ee
Here, the second term on the LHS stems from the expression
$\bra{-\D^2\sigma_{ab}}\tilde\bb^b$ which emerges during the
calculation. As expected, equation \reff{eq:ddotI} is equivalent
to multiplying the $m$-mode shear equation \reff{sigmadot} with
the magnetic field $\tilde\bb^a$. We remind the reader that the
derivation of equation \reff{eq:ddotI} does not rely on the MHD
approximation and is indeed valid for all values of the curvature
index $K$ and is also independent of the equation of state.
\subsection{Governing equation for $\ff^a$}\label{sec:F}
In the ideal MHD limit, the electric field is expressed as the
cross-product of the primary magnetic field and the velocity,
$E_a=-\epsilon_{abc}v^b\tilde B^c$. Using the linear velocity
propagation equation \reff{dotv} together with the linear
evolution equation \reff{ind} for the magnetic field, one easily
finds $\dot E_{<a>}+\Theta E_a=0$, and therefore
\ber
\bra{\curl E_a}^{~\dot{}}_{\perp} &=& \curl \dot
E_{<a>}-\sfrac13\Theta\curl E_a \nonumber \\
&=& -\sfrac43 \Theta \curl E_a\;.
\eer
This further implies that the evolution equation of the MHD
contribution encoded in $\ff^a$ is simply given by
\be
{\dot\ff}_{<a>}+\sfrac23\Theta\ff_{a} = 0\;. \label{eq:dotFa}
\ee
It follows that the term $\ff_a$ evolves like the electric field
$E_a$, and decays as $\sim a^{-2}$.
\section{Solutions for spatially flat Universes}\label{sec:phase4}
Because the derived MHD equations above are only valid in the cold
plasma limit, we only investigate the dust solutions for
spatially flat models with zero cosmological constant. It is
convenient to use the dimensionless time variable $\tau \equiv
\sfrac32H_0\bra{t-t_0}+1$ introduced in \cite{us}. In terms of
this variable, the Hubble parameter evolves simply as $H_0/\tau$
and the scale factor obeys $a=a_0\tau^{2/3}$, where the zero index
indicates evaluation at some arbitrary initial time $t_0$.

Employing the time variable $\tau$, the equation for the
interaction variable \reff{eq:ddotI} may be written as 
\be
\frac94\,\ii^{~\prime\prime}_{<a>}+\frac{15}{2\tau}\,\ii^{~\prime}_{<a>}
+\bras{\frac{3}{2\tau^2}+\bra{\frac{m}{a_0H_0}}^2\tau^{-
\frac{4}{3}}}\ii_{<a>}=0\;,\label{intinhomo}
\ee
while the equation \reff{eq:dotFa} for the MHD term transforms into
\be
{\ff}_{<a>}^{~\prime}+\frac{4}{3\tau}\ff_{a} = 0\;, \label{dotFa}
\ee
whose solution is simply
\be
\ff_a=\ff_a^0\tau^{-\frac43}=\ff_a^0(a/a_0)^{-2}.\label{f}
\ee
The generated magnetic field will typically depend upon $x\equiv
m/(a_0H_0)=2\pi(\lambda_{\rm H}/\lambda_{\tilde
B})_0=2\pi(\lambda_{\rm H}/\lambda_{\rm GW})_0$, which is the
ratio between the size of the magnetised field region and the
horizon size when the interaction begins.
\subsection{Limiting case where $x\rightarrow 0$ }\label{longinhomo}
If the value of $x$ is so small that we can drop the term $x^2$ in
equation \reff{intinhomo}, the general solution for the interaction
variable is
\be
\ii_{a}(\tau)=C_1
\tau^{-\frac{1}{3}}+C_2\tau^{-2}\;,\label{longinteract}
\ee
where $C_1$ and $C_2$ are integration constants. Since for the
generation of magnetic fields the dominant mode is more important,
we set $C_2=0$ and obtain the solution
\be
\ii_a(\tau)=\ii_a^0\tau^{-\frac{1}{3}}=\sigma^0_{ab}\tilde
B_0^b\tau^{-\frac{1}{3}}\;. \label{x0}
\ee
It is now a very simple exercise to integrate the induction
equation \reff{beta2} to find
\be
\sfrac32H_0\bb_a = \sfrac32\ii_a^0\bra{\tau^{\frac23}-1}
-\ff_a^0\bra{\tau^{-\frac13}-1}\;,\label{ind1}
\ee
where the integration constant was determined by requiring the
generated magnetic field to vanish initially. It follows that
the total magnetic field measured by the fundamental observer
due to the zero-zero mode interaction becomes
\be \label{llloo}
B^{(0-0)}(a) = \tilde
B_0^{(0)}\bra{\frac{a_0}{a}}^2\bras{1+\frac{\sigma_0^{(0)}}{H_0}
\brac{\frac{a}{a_0}-1} - \frac23\,\frac{\ff_0^{(0)}}{H_0\tilde
B_0^{(0)}}\brac{\bra{\frac{a}{a_0}}^{-\frac12}-1} }\;,
\ee
where the second term in the square bracket originates from the
interaction of the magnetic field $\tilde B_a$ with the shear
(already obtained in \cite{us}), while the third term represents the
MHD contribution, which is the interaction of the magnetic field
$\tilde B_a$ with the plasma velocity perturbation $v_a$. Note that
the MHD contribution slowly decays away as the scale factor
increases, in contrast to the gravito-magnetic part, which linearly
grows with the scale factor. However, in the long-wavelength limit
the contribution due to the GWs is negligible since one typically
finds the shear anisotropy to be very small, $\bra{\sigma/H}_0 \ll
1$. Moreover, since we may approximate $\mathcal F_0 = F_0\approx
(v\tilde B/\lambda_{\rm\tilde B})_0$ due to $F_a=-\curl E_a$ and the
MHD relation \reff{Einhomo}, we see that the factor $\mathcal
F_0/(H_0\tilde B_0)$ in \reff{llloo} is proportional to $x \ll 1$,
and the MHD contribution turns also out to be negligible. Consequently,
in the long-wavelength limit, there is no amplification of the initial
magnetic field by GWs or velocity perturbations and one needs to consider 
the general case in order to look for an amplification.
\subsection{General case with $x\neq 0$ }
When $x$ is not negligible (the magnetised region is strictly
finite), the general solution to the equation \reff{intinhomo} is
found to be
\be
\ii_{a}(\tau)=\tau^{-\frac{7}{6}}
\bras{D_1\,J_1\bra{\sfrac{5}{2},2\,x\,
\tau^{\frac{1}{3}}}+D_2\,J_2\bra{\sfrac{5}{2},2\,x\;,
\tau^{\frac{1}{3}}}}\;,\label{intsolgen2}
\ee
where $D_1$, $D_2$ are integration constants and $J_1$, $J_2$
denote Bessel functions of the first and second kind,
respectively. Since we are only interested in the dominant
contribution, we set $D_2=0$ as before, noting that the Bessel
function of the second kind is decaying on super-horizon scales
$x\ll 1$. The remaining integration constant takes then the value
\be
D_1 =
\frac{\sqrt{\pi}x^{\frac52}}{4x^2\sin(2x)
-3\sin(2x)+6x\cos(2x)}\,\sigma_0\tilde
B_0\;.
\ee
Assuming that the induced magnetic field is zero initially when
the interaction begins $(\tau=1)$, the solution for the rescaled
magnetic field then becomes
\be
\bb(\tau) = \frac{\sigma_0\tilde
B_0}{H_0y}\bras{\sfrac12\sin(2x)-x\cos(2x) +
x\cos(2x\tau^{\frac13})\tau^{-\frac23}-
\sfrac12\sin(2x\tau^{\frac13})\tau^{-1}}
-\frac23\frac{F_0}{H_0}\bra{\tau^{-\frac13}-1}\;,\label{Da}
\ee
where we defined $y \equiv 4x^2\sin(2x)-3\sin(2x)+6x\cos(2x)$ and
made use of \reff{dotFa}. Notice that in the limit $x\rightarrow
0$ one recovers the result \reff{ind1}. Had we instead focused on
the other branch of the solution \reff{intsolgen2}, we would find
\reff{Da} again but with the $\sin$ and $\cos$
functions as well as some signs interchanged. Hence, the
$m-m$ mode interaction generated magnetic field as seen by the
fundamental observer moving with 4-velocity $u^a$ is given by the
expression
\be
B^{(m-m)}(a) = \tilde B_0^{(m)}\bra{\frac{a_0}{a}}^2\bras{
\frac{\sigma_0^{(m)}}{H_0}\;,
\frac{\sfrac12\sin(2x)-x\cos(2x)}{4x^2\sin(2x)-3\sin(2x)+6x\cos(2x)}
+ \frac23\frac{F_0^{(m)}}{H_0\tilde B_0^{(m)}}
+\mathcal{O}(a^{-\frac12})}\;;
\ee
here, the non-displayed terms are decaying with time and therefore
irrelevant for the amplification process but can be inferred
easily from \reff{Da} if required.

What happens if we take the full solution \reff{intsolgen2} of the
interaction variable into account instead of just looking at one
branch? Since both the Bessel functions of the first as well as
the second kind are merely $\sin$ and $\cos$ functions modified by
the \emph{same} damping envelopes, one should in principle
consider both branches even though their asymptotic behaviour is
different. In this case, the integration constants $D_1$ and $D_2$
are determined by requiring that initially one has
$\ii_0=\ii(\tau=1)=\sigma_0^{(m)}\tilde B_0^{(m)}$ together with
$\ii^{~\prime}_0=\sigma_0^{\,\prime(m)}\tilde B_0^{(m)}$. The
exact solution for the rescaled magnetic field is then found in
analogy with the above example [cf. \reff{Da}], but it is too large
to display it in full. However, the dominant contribution to the 
generated magnetic field can be written down in simple terms as
\be
B^{(m-m)}(a) = \tilde B_0^{(m)} \bra{\frac{a_0}{a}}^2
\bras{\frac{3}{2x^2}\frac{2\sigma_0^{(m)}+\sigma_0^{\,\prime(m)}}{H_0}
+ \frac23\frac{F_0^{(m)}}{H_0\tilde B_0^{(m)}}
+\mathcal{O}(a^{-\frac12})}\;,
\ee
where it was again assumed that there is no generated magnetic
field initially. If we employ the natural length scale
$\lambda_{\rm \tilde B}$, inherent to the problem under
consideration, we may estimate $\dot\sigma_0 \approx
(\sigma/\lambda_{\rm\tilde B})_0$, implying $\sigma_0^{~\prime}
\approx 2/3(\sigma/H_0\lambda_{\rm\tilde B})_0$, and also
$F_0\approx (v\tilde B/\lambda_{\rm\tilde B})_0$ due to
$F_a=-\curl E_a$ and the MHD relation \reff{Einhomo}. Remembering
further the definition of $x=2\pi(\lambda_{\rm H}/\lambda_{\tilde
B})_0$, we can finally write down the expression for the
\emph{total} magnetic field in a more convenient form:
\be
\label{main}
B^{(m-m)}(a) = \tilde B_0^{(m)} \bra{\frac{a_0}{a}}^2
\bras{1+\frac{3}{4\pi^2}\bra{\frac{\lambda_{\rm\tilde
B}}{\lambda_{\rm H}}}^2_0
\frac{\sigma_0^{(m)}}{H_0}\brac{1+\frac{1}{3}\bra{\frac{\lambda_{\rm
H}}{\lambda_{\rm\tilde B}}}_0} + \frac23 v_0
\bra{\frac{\lambda_{\rm H}}{\lambda_{\rm\tilde B}}}_0 +
\mathcal{O}(a^{-\frac12}) }\;.
\ee
This is our main result - it shows in detail how the magnetic field,
resulting from the interaction of the background magnetic field
$\tilde B_0$ with GWs and velocity perturbations $v_0$ in the
plasma, depends on the initial conditions.

Observe that at super-Hubble scales the MHD contribution becomes
completely negligible mirroring the observation that plasma effects 
are typically more important on small scales. Our main result 
directly generalises our previous result (49) in \cite{us} derived 
for the case of a homogeneous magnetic field to the inhomogeneous case. 
It should be stressed that the use of ideal MHD in the cold plasma 
limit allowed for a self-consistent treatment of the electric fields 
and plasma currents.
\section{Discussion}
If we look only at \emph{super-horizon} scales and divide the result
\reff{main} through the energy density of the background radiation,
$\mu_{\gamma}$, (which decays in the same manner as the original
magnetic field), the dominant contribution is then given by
\be
\frac{B}{\mu^{1/2}_{\gamma}}
\simeq
   \bras{1+\frac{1}{10}
\bra{\frac{\lambda_{\tilde \mathrm{B}}}{\lambda_{\mathrm{H}}}}^2_0
\bra{\frac{\sigma}{H}}_0}\bra{\frac{\tilde B}{\mu^{1/2}_{\gamma}}}_0\;,
\label{summary2}
\ee
where the wavenumber indices have been suppressed and the zero
suffix indicates the time when the interaction begins. This result
\reff{summary2} was already found in Betschart \emph{et al.}
\cite{us} and previously reported by Tsagas \emph{et al.}
\cite{GWamp}, a paper which employed the weak field approximation.
The result \reff{summary2} can be applied to the reheating phase of
the  Universe at the end of inflation, for which the effective
equation of state was that of dust (cf. \cite{us,GWamp} for an
application).

On the other hand, at sub-horizon scales  the main part of the
magnetic field is given by
\be
\frac{B}{\mu^{1/2}_{\gamma}} \simeq \bras{1+\frac{2}{3}\,v_0
\bra{\frac{\lambda_{\mathrm{H}}}{\lambda_{\tilde
\mathrm{B}}}}_0}\bra{\frac{\tilde B}{\mu^{1/2}_{\gamma}}}_0 ,
\label{summary3}
\ee
which could be applied to the matter-dominated phase of the
Universe. In order to obtain an order-of-magnitude estimate we
assume that the velocity perturbations (resulting from Thomson
scattering) in the primordial plasma in effect start to interact
with the pre-existing magnetic field $\tilde B$ somewhat after
matter-radiation equality with a redshift of roughly $(z_{\textrm
{eq}} \simeq 10^4)$. The horizon at matter-radiation decoupling was
$\lambda_{\mathrm
H}^{\mathrm{eq}}=H_0^{-1}(1+z_{\mathrm{eq}})^{-3/2}\simeq10^{-2}
\,\mathrm{Mpc}$, where $H_0$ denotes today's Hubble constant. A
typical size of a seed field required for the dynamo mechanism is
$\lambda_{\tilde B} \simeq 10\,\mathrm{kpc}$ on a comoving scale
today, hence
$\lambda_{\tilde\mathrm{B}}^{\mathrm{eq}}=(1+z_{\mathrm{eq}})^{-1}\lambda_{\tilde
B}\simeq 1\,\mathrm{pc}$. From cosmic microwave background (CMB)
measurements we know that at decoupling $(z_{\mathrm{dec}}\simeq
10^3)$ the velocity perturbations satisfy $v_{\mathrm{dec}}\leq
10^{-5}$, while we read off from equation \reff{dotv} that they
decay like $a^{-1}$. It follows that $v_{\mathrm{dec}} =
v_{\mathrm{eq}}(a_{\mathrm{eq}}/a) = v_{\mathrm{eq}}\tau^{-2/3} =
(1+z_{\mathrm{dec}})(1+z_{\mathrm{eq}})^{-1}v_{\mathrm{eq}}
\simeq10^{-1}\,v_{\mathrm{eq}}$, and whence $v_{\mathrm{eq}}\simeq
10^{-4}$. Taking everything together, one obtains  the boost factor
in \reff{summary3} to be of order unity, that is,
$\bra{v\lambda_\mathrm{H}/\lambda_{\tilde
\mathrm{B}}}_{\mathrm{eq}}\simeq 1$. Given that we can rely on the
MHD approximation at the early stages of the matter-dominated
scenario, the velocity perturbations in the plasma will lead at best
to a doubling of the initial magnetic field strength.

Comparing the  result \reff{summary2} with the final solution
presented in Betschart \emph{et al.} indicates that the magnitude of
the amplification due to the interaction between GWs and magnetic
fields is proportional to the square of the ratio of the coherence
length $\lambda_{\tilde \mathrm{B}}$ of the initial magnetic field
and the initial size of the horizon $\lambda_{\mathrm{H}}$ in both
the homogeneous and inhomogeneous magnetic field cases. The
additional MHD part of the field in \reff{main} above arises from
the forcing term $F^a \equiv  -\curl E_a = 2 \D^b\bra{v_{[a}B_{b]}}$
whose time behaviour is obtained from a first-order propagation
equation in which a Laplacian does not appear, its general solution
therefore being independent of wavenumber and scale (of course, the
initial conditions still depend on the size of the interaction
region). It was found that the seed field's interaction with GWs is
only important at super-horizon scales, while the interaction with
plasma velocity perturbations dominates at sub-horizon scales. It is
worth pointing out that there is no amplification at all in the
long-wavelength limit.

In contrast with the results in Betschart \emph{et al.} \cite{us},
the generated magnetic field modes now have wavenumbers that are
constructed from those of the interacting field and GWs and thus
differ from those of the GWs alone. If the GW has wave vector
$k_a$ and the `background' magnetic field a wave vector $m_a$ (where
$\lambda_{\tilde \mathrm{B}}=2\pi a /m$ corresponds to the size of
the magnetic inhomogeneity), the wavenumber $\ell$ of the induced
field satisfies $\ell^2 = (k_a + m_a)(k^a + m^a) = k^2 + m^2 + 2m_a
k^a$. If for simplicity the first order wave vectors are assumed to
be orthogonal, $k^a m_a = 0$, then the $m-m$ interaction yields a
field with wavenumber $\ell^2 = 2 m^2$ ($4 m^2$ in the parallel
case). It follows that the characteristic wavelength $\lambda = 2
\pi a/ \ell$ of the induced field is somewhat shorter than the
original B-field by the superposition of the corresponding magnetic
and GW wavenumbers.

It is worth noting that the presence of a spatially
homogeneous field is only consistent in the flat Universe \cite{us},
whereas no restrictions on the spatial geometry arise at any stage in 
the
derivation of the evolution equations with $(\D^b B^a)\neq 0$ in this
paper. This makes sense when we consider the spectrum of allowed 
wavenumbers
for different geometries.  For open models ($K = +1$), the lowest
wavenumber is $n=3$. For closed models ($K = -1$), we find that $n\geq
1$. However,
the spectrum of wavenumbers in a flat spacetime ($K = 0$) is
continuous, with $n\geq 0$. Given that the spatial homogeneity of
the background field in \cite{us} restricts its associated wavenumber
to $n=0$, we see that this eigenvalue can only be accommodated in
a flat Universe.

More interesting is the relationship between curvature and magnetic
fields. Einstein's theory is geometrical which implies that vectors
are directly coupled to the spacetime curvature via the Ricci
identity \cite{christos2}.  In \cite{TM}, Tsagas and Maartens find
that the evolution equation of the spatial gradient of the magnetic
field contains a term $\epsilon_{acd}B^c H^{~d}_b\sim {\mathcal
O}(\epsilon_B\epsilon_g)$, which alludes to a non-local coupling to
curvature. This result is confirmed in \cite{TB} with the appearance
of a term containing the spatially projected Riemann curvature
tensor in the propagation equation of $\bra{\D_a B_b}$ and indicates
that the curvature sources magnetic inhomogeneities. In the case of
a homogeneous magnetic field in Betschart \textit{et al.}, the
spatial gradients of the magnetic field are at least second-order.
In order to preserve the spatial uniformity of the first-order field
through time, the spatially projected Riemann tensor may have to
vanish to prevent it from sourcing $(\D_a B_b)$ so that the spatial
gradients remain small and continue to contribute at higher-order
only.  In the analysis of the inhomogeneous magnetic field presented 
here,
magnetic spatial eddies exist at first-order and for this reason, the 
boost
from the coupling between the field and the curvature need not
necessarily be eliminated. In fact, such couplings are explicitly
retained in our approach via the standard commutation relations
(cf. the appendix).
\section{Conclusion}
Although the focus of magnetogenesis in recent years has been the
generation of \emph{large-scale} magnetic fields, a self-sufficient
mechanism still evades us. The galactic dynamo is indeed physically
feasible and has been shown to generate fields with strengths
matching current observations, but requires a reasonably strong seed
field to work.  To make this theory more robust, we need to find a
way of producing seed fields that are suitable for subsequent
amplification by the dynamo. Cosmological perturbations have been
identified as a possible source of primordial magnetic field
amplification which is present in the pre-recombination era.
Considering second-order couplings between electrons, photons
and protons, the electric current induced by the plasma
vorticity (a known source of magnetic fields) and the
additional contribution of the photon anisotropic stress are
found to yield a magnetic field that is a
sufficient seed for the dynamo to work \cite{Tak}. The coupling of
density and velocity perturbations that are naturally occurring in
the early Universe are shown to lead to similar resultant fields in
the context of a relativistic charged multi-fluid \cite{mark}.

In this paper we built on previous work in Betschart \emph{et al.},
in which the full set of equations determining the evolution of the
gravitational waves and the generated electromagnetic fields was
presented, initially for the case of a homogeneous magnetic field,
and generalised the analysis to the case of a spatially
inhomogeneous magnetic field using the magnetohydrodynamic
approximation (restricting ourselves to the dust case).  Analysing
the equations for a spatially flat dust FLRW Universe, we were able
to confirm our previous results. In particular, we do not find any
amplification in the case of the long-wavelength limit.

Over and above the presentation of a physically viable mechanism for
primordial magnetogenesis, this paper also establishes a formalism
which guides the choice of proper second-order gauge-invariant
variables. Using this methodology, one is able to obtain results
in terms of clearly defined quantities, with no ambiguity concerning
the physical validity of the variables.

The possibility of obtaining seed magnetic fields of sufficient
magnitude via a combination of inflationary physics and standard MHD
theory is an exciting prospect. However, many of the parameters
involved in estimating the size of the effect are not well known.
For instance the spectrum of gravitational waves predicted by
inflationary theory still has to be verified, as well as the
large-scale structure of cosmological magnetic fields. For this
purpose, more precise and complete measurements (such as the Planck
mission) of the, e.g. polarization of the CMB would give much needed
information \cite{Ensslin-etal}. Moreover, studies of interactions 
between incoherent gravitational wave distributions and turbulent magnetic
fields could also be done in order to obtain a more detailed picture.
This is left for future research.


\acknowledgments

CZ, GB and PKSD thank the Physics Department of Ume{\aa} University
for hospitality while part of this work was carried out. This
research was supported by a Sida/NRF grant, and was partially supported by
the Swedish Research Council.
CZ acknowledges support from a Domus A scholarship awarded
by Merton College. GB is funded by a Lady
Davis Postdoctoral Fellowship

\appendix*
\section{Commutation relations}
Here we present various commutator relations which have been used
in the text. The relations are given up to second
order in our perturbation scheme. The vanishing of vorticity,
$\omega_{ab}=0$, is assumed throughout in conjunction with the
constraints $\D_a\mu=\D_a p=0$ which isolate the pure tensor
modes. All appearing tensors are PSTF, $S_{ab}=S_{<ab>}$, and all
vectors $V_a,\,W_a$ are purely spatial.

Commutators for first-order vectors $V_a$:
\ber
\bra{\D_a V_b}^{\dot{}}_{\perp} &=& \D_a
\dot{V}_b-\sfrac{1}{3}\Theta
\D_a V_b -\sigma_a^{~c}\D_cV_b+H_a^{~d}\,\epsilon_{dbc}\,V^c \\
\bra{\curl V_a}^{\dot{}}_{\perp} &=&
\curl\dot{V}_a-\sfrac{1}{3}\Theta\,
\curl V_a -\sigma_a^{~b}\,\curl V_b -H_{ab}\,V^b \\
\D_{[a}\,\D_{b]} V_c &=&
\bras{\sfrac{1}{9}\Theta^2-
\sfrac{1}{3}\bra{\mu+\Lambda}}V_{[a}\,h_{b]c}+\bra{\sfrac{1}{3}\Theta\,\
sigma_{c[a}-E_{c[a}}V_{b]}
\nonumber\\&&+h_{c[a}\bra{E_{b]d}-\sfrac{1}{3}\Theta\,\sigma_{b]d}}
V^d
\eer
Commutators for first-order tensors $S_{ab}$:
\ber
\bra{\D_a S_{bc}}^{\dot{}}_{\perp} &=& \D_a
\dot{S}_{bc}-\sfrac{1}{3}\Theta \D_a S_{bc}-\sigma_a^{~d} \D_d
S_{bc}+2H_a^{~d}\,\epsilon_{de(b}\,S_{c)}^{~~e}
\\
\bra{\D^b S_{ab}}^{\dot{}}_{\perp} &=& \D^b
\dot{S}_{ab}-\sfrac{1}{3}\Theta \D^b S_{ab}-\sigma^{bc}\D_c
S_{ab}+\epsilon_{abc}\,H^b_{~d}\, S^{cd}
\\
   \bra{\curl S_{ab}}^{\dot{}}_{\perp}&=&\curl
\dot{S}_{ab}-\sfrac{1}{3}\Theta\, \curl S_{ab} -\sigma_e^{~c}\,
\epsilon_{cd(a}\,D^e S_{b)}^{~~d} +3 H_{c<a}S_{b>}^{~~c}
\\
\curl\curl S_{ab}&=&-D^2 S_{ab}+\bra{\mu+\Lambda
-\sfrac{1}{3}\Theta^2}S_{ab} +\sfrac{3}{2}D_{<a}D^c
S_{b>c}\nonumber\\
&&+3 S_{c<a}\bra{E_{b>}^{~~c}-\sfrac{1}{3}\Theta\sigma_{b>}^{~~c}}
\eer
Commutators for second-order vectors $W_a$:
\ber
\bra{\D_a W_b}^{\dot{}}_{\perp}&=&\D_a
\dot{W}_b-\sfrac{1}{3}\Theta\, \D_a W_b
\\
\D_{[a}\,\D_{b]} W_c &=&
\bras{\sfrac{1}{9}\Theta^2-\sfrac{1}{3}\bra{\mu+\Lambda}}W_{[a}\,h_{b]c}
\\
\curl\curl W_a &=& -\D^2W_a +\D_a \bra{\di W}
+\sfrac23\bra{\mu+\Lambda -\sfrac{1}{3}\Theta^2}W_{a}
\eer


\end{document}